\documentclass[10pt]{llncs}
\usepackage{booktabs} 
\usepackage{array, makecell} 
\usepackage{textcomp} 
\usepackage{paralist} 
\usepackage[font=small]{caption}
\usepackage{amsmath}
\usepackage{amsfonts} 
\usepackage{amssymb} 
\usepackage{pifont} 
\newcommand{\cmark}{\ding{51}} 
\newcommand{\xmark}{\ding{55}} 
\usepackage{graphicx}
\usepackage{url}
\usepackage{color} 
\usepackage[ruled,vlined]{algorithm2e} 
\newcommand{\quotes}[1]{``#1''} 
\newcommand{\name}{DSTC\xspace}
\newcommand{\namelong}{DNS-based Strict TLS Configurations\xspace}

\usepackage{msc} 
\usepackage{xcolor}
\usepackage{multicol} 
\usepackage{multirow} 
\usepackage[misc]{ifsym} 
\usepackage{hyperref}

\newcommand*\justify{
  \fontdimen2\font=0.4em 
  \fontdimen3\font=0.2em 
  \fontdimen4\font=0.1em 
  \fontdimen7\font=0.1em 
  \hyphenchar\font=`\- 
}

\begin{document}
\title{\name: \namelong}

\author{Eman Salem Alashwali\inst{1,}\inst{2}\textsuperscript{(\Letter)} \and Pawel Szalachowski\inst{3}} 
\institute{University of Oxford, United Kingdom \\
\and King Abdulaziz University (KAU), Saudi Arabia \\
\email{eman.alashwali@cs.ox.ac.uk} \\
\and Singapore University of Technology and Design (SUTD), Singapore\\
\email{pawel@sutd.edu.sg}
}

\maketitle
\setcounter{footnote}{0}

\begin{abstract}
Most TLS clients such as modern web browsers enforce coarse-grained TLS security configurations. They support legacy versions of the protocol that have known design weaknesses, and weak ciphersuites that provide fewer security guarantees (e.g. non Forward-Secrecy), mainly to provide backward compatibility. This opens doors to downgrade attacks, as is the case of the POODLE attack \cite{moller14}, which exploits the client's silent fallback to downgrade the protocol version to exploit the legacy version's flaws. To achieve a better balance between security and backward compatibility, we propose a DNS-based mechanism that enables TLS servers to advertise their support for the latest version of the protocol and strong ciphersuites (that provide Forward-Secrecy and Authenticated-Encryption simultaneously). This enables clients to consider prior knowledge about the servers' TLS configurations to enforce a fine-grained TLS configurations policy. That is, the client enforces \textit{strict} TLS configurations for connections going to the advertising servers, while enforcing \textit{default} configurations for the rest of the connections. We implement and evaluate the proposed mechanism and show that it is feasible, and incurs minimal overhead. Furthermore, we conduct a TLS scan for the top 10,000 most visited websites globally, and show that most of the websites can benefit from our mechanism.  
\end{abstract}

\section{Introduction}
\label{sec:intro}
Websites\footnote{Throughout the paper we use the terms website, server, and domain, interchangeably to refer to an entity that offers a service or content on
the Internet.} vary in the sensitivity of the content they serve and in the
level of communication security they require. For example, a connection to an
e-banking website to make a financial transaction carries more sensitive data
than a connection to an ordinary website to view public news. A close look at
how mainstream TLS clients (e.g. web browsers) treat these differences reveals
that they enforce coarse-grained TLS
security configurations, i.e. a \quotes{one-size-fits-all} policy. They\footnote{We tested the following browsers: \texttt{Google Chrome} version 67.0.3396.87, \texttt{Mozilla Firefox} version 60.0.2, \texttt{Microsoft Internet Explorer} version 11.112.17134.0, \texttt{Microsoft Edge} version 42.17134.1.0, and \texttt{Opera} version 53.0.2907.99.} support legacy versions of the protocol that have known design weaknesses and weak ciphersuites that provide fewer security guarantees, e.g. non Forward-Secrecy (non-FS), and non Authenticated-Encryption (non-AE), mainly for backward compatibility. \par

Supporting legacy versions or weak ciphersuites provides backward compatibility, but opens doors to downgrade attacks. In downgrade attacks, an active Man-in-the-Middle (MitM) attacker forces the communicating parties to operate in a mode weaker than they both support and prefer. Several studies illustrate the practicality of
downgrade attacks in TLS~\cite{adrian15,aviram16,beurdouche15,beurdouche14,bhargavan2016downgrade,bhargavan2016transcript,moller14}. Despite numerous efforts to mitigate them, they continue to appear up until 2016 in a draft for the latest version of TLS, TLS~1.3 \cite{bhargavan2016downgrade}. Previous attacks have exploited not only design vulnerabilities, but also implementation and trust model vulnerabilities that bypass design-level mitigations such as the handshake messages (transcript) authentication. For example, the POODLE \cite{moller14}, DROWN \cite{aviram16}, and \texttt{ClientHello} fragmentation \cite{beurdouche14} downgrade attacks. \par

Clearly, disabling legacy TLS versions and weak ciphersuites at both ends prevents downgrade attacks: There is no choice but the latest
version and strong ciphersuites. However, the global and heterogeneous nature of the
Internet have led both parties (TLS client vendors and server
administrators) to compromise some level of security for backward compatibility.
Furthermore, from a website perspective, supporting legacy TLS versions and weak ciphersuites may
not only be a technical decision, but also a business decision not to lose customers for another website. \par

However, we observe that if the client has prior knowledge about the servers' TLS configurations, a better balance between security and backward compatibility can be achieved, which reduces the downgrade attack's surface. Given prior knowledge about the servers' ability to meet the latest version of the protocol and strong ciphersuites, the client can change its behaviour and enforce a \textit{strict} TLS configurations policy when connecting to these advertising servers. \par 

In this paper, we try to answer the following question: \textbf{\textit{How to enable domain owners to advertise their support for the latest version of the TLS protocol and strong ciphersuites to clients in a usable and authenticated manner? This is in order to enable clients to make an informed decision on whether to enforce a \textit{strict} or \textit{default} TLS configurations policy before connecting to a server. }}\par

Our contributions are as follows: First, we propose a mechanism that enables domain owners to advertise their support for the latest version of the TLS protocol and strong ciphersuites. This enables clients to enforce \textit{strict} TLS configurations when connecting to the advertising domains while enforcing \textit{default} configurations for the rest of the domains. We show how our mechanism augments clients' security to detect certain types of downgrade attacks and server misconfiguration. Second, we implement and evaluate a proof-of-concept for the proposed mechanism. Finally, we examine the applicability of our mechanism in real-world deployment by conducting a TLS scan for the top 10,000 most visited websites globally on the Internet. \par

\section{Background}
\label{sec:background}
\subsection{Domain Name System (DNS)}
Domain Name System (DNS)~\cite{mockapetris1987domain} is a decentralized and
hierarchical naming system that stores and manages information about domains.
DNS introduces different types of information which are stored in dedicated
resource records.  For example, the \texttt{A} resource records are used to point a
domain name to an IPv4 address, while \texttt{TXT} records are
introduced for storing arbitrary human-readable textual information. DNS is primarily used for resolving domain names to IP addresses, and usually
this process precedes the communication between hosts. Whenever a client wants
to find an IP address of a domain, for example \quotes{www.example.com}, it
contacts the DNS infrastructure that resolves this name recursively. Namely,
first, a DNS root server is contacted to localize an authoritative server for
\quotes{com}, then this server helps to localize \quotes{example.com}'s authoritative
server, which at the end returns the address of the target domain. To make this
process more efficient, the DNS infrastructure employs different caching
strategies.

\subsection{Domain Name System Security Extension (DNSSEC)}
DNS itself does not provide (and was never designed to provide) any protection
of the resource records returned to clients. DNS responses can be freely manipulated
by MitM attackers. DNS Security Extensions
(DNSSEC) \cite{larson2005dns} is an extension of DNS which aims to improve this
state. DNSSEC protects DNS records by adding cryptographic signatures to assert
their origin authentication.  In DNSSEC, each DNS zone has its Zone Signing Key
(\texttt{ZSK}) pair. The \texttt{ZSK}'s private-key is used to sign the DNS records.
Signatures are published in DNS via dedicated \texttt{RRSIG} resource
records. The \texttt{ZSK} public-key is also published in DNS in the special
\texttt{DNSKEY} record. The \texttt{DNSKEY} record is also signed with the
private-key of a Key Signing Key (\texttt{KSK}) pair, which is signed by an upper-level \texttt{ZSK} (forming a trust chain). To validate authentication of the DNS received responses,
clients have to follow the trust chain till the root.

\subsection{Transport Layer Security (TLS)}
Transport Layer Security (TLS) is one of the most important and widely-deployed client-server protocols that
provides confidentiality and data integrity on the Internet. It was formerly known as the Secure Socket Layer (SSL). TLS consists of
multiple sub-protocols including the TLS handshake protocol that is
used for establishing TLS connections. A particularly important and security-sensitive aspect of the handshake
is the selection of the protocol version and the cryptographic algorithms with their
parameters (i.e. ciphersuites). Every new version of TLS prevents security attacks in previous versions. Some ciphersuites provide more security guarantees than others. For example, Forward Secrecy (FS) is a property that guarantees that a compromised long-term private key does not compromise past session keys \cite{menezes96}. Both finite-field Ephemeral Diffie-Hellman (\texttt{DHE}) and Elliptic-Curve Diffie-Hellman (\texttt{ECDHE}) key-exchange algorithms provide the FS property. On the other hand, RSA does not provide this property. Similarly, Authenticated Encryption (AE) provides confidentiality, integrity, and authenticity simultaneously such that they are resilient against padding oracle attacks \cite{cloudflare16}\cite{vaudenay02}. \texttt{GCM}, \texttt{CCM}, and \texttt{ChaCha-Poly1305} ciphers provide the AE property while the \texttt{CBC} MAC-then-Encrypt ciphers do not provide authentication and encryption simultaneously, and hence do not provide the AE property.   
 
\subsection{TLS Version and Ciphersuite Negotiation} \label{tls-negotiation}
We base our description on TLS~1.2 \cite{rescorla08tls12}. The coming version TLS~1.3 is still a draft \cite{rescorla18tls13}. At the beginning of a new TLS handshake, the client sends a \texttt{ClientHello} (\texttt{CH}) message to the server. The \texttt{ClientHello} contains several parameters including the supported versions and ciphersuites. In TLS~1.2 the client sends its supported versions as a single value which is the maximum supported version by the client $vmax_C$, while in TLS~1.3, they are sent as a list of supported versions [$v_1,...,v_n$] in the \texttt{supported\_versions} extension. The $vmax_C$ is still included in TLS~1.3 \texttt{ClientHello} for backward compatibility and its value is set to TLS~1.2. The \texttt{supported\_versions} extension is not for pre TLS~1.3 versions \cite{rescorla18tls13}. The client's supported ciphersuites are sent as a list [$a_1,...,a_n$]. Upon receiving a \texttt{ClientHello}, the server selects the version and ciphersuite that will be used in that session, and responds with a \texttt{ServerHello} (\texttt{SH}) containing the selected version $v_S$ and the selected ciphersuite $a_S$. Ideally, these two values are influenced by the client's offered versions and ciphersuites. If the server selected a version lower than the client's maximum version, most TLS clients fall back silently to the lower versions (up to TLS~1.0 in all mainstream browsers today). The silent fallback mechanism can be abused by attackers to perform downgrade attacks as shown in the POODLE \cite{moller14}, a variant of DROWN \cite{aviram16}, and ClientHello fragmentation \cite{beurdouche14} downgrade attacks. 

\subsection{TLS Downgrade Attacks}
\label{sec:back:tls_downgrade}
\begin{figure}[!tp]
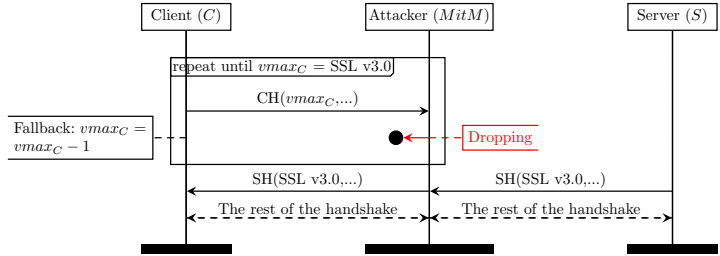
 
	\centering
	\resizebox{0.80\textwidth}{!}{
		\setmsckeyword{} 
		\drawframe{no} 
		\begin{msc}[large values, /msc/level height=0.6cm, /msc/label distance=0.5ex , /msc/first level height=0.6cm, /msc/last level height=0.6cm, /msc/top head dist=0, /msc/bottom foot dist=0, /msc/environment distance=0, /msc/foot height=1.5ex]{}
			\setlength{\instwidth}{2\mscunit} 
			\setlength{\instdist}{3\mscunit} 
			\declinst{I}{}{Client ($C$)}%
			\declinst{M}{}{Attacker ($MitM$)}%
			\declinst{R}{}{Server ($S$)}%
			\inlinestart[/msc/left inline overlap=0.35cm, /msc/right inline overlap=0.35cm]{exp1}{repeat until $vmax_{C}$ = SSL~v3.0}{I}{M}
			\nextlevel
			\nextlevel
			\mess {CH($vmax_{C}$,...)}{I}{M}
			\nextlevel
			\lost[side=left, color=red]{}{}{M}
			\msccomment[msccomment distance=0.75cm, side=right, color=red]{Dropping}{M}
			\msccomment[msccomment distance=0.75cm, /msc/every msccomment/.append style={text width=3cm}]{Fallback: $vmax_{C}=vmax_{C}-1$}{I}
			\nextlevel
			\inlineend{exp1}
			\nextlevel
			\mess {SH(SSL~v3.0,...)} {R}{M}
			\mess {SH(SSL~v3.0,...)} {M}{I}
			\nextlevel
			\mess*{The rest of the handshake}{I}{M}
			\mess*{The rest of the handshake}{M}{R}
			\mess*{}{R}{M}
			\mess*{}{M}{I}
		\end{msc}
	} 
	\vspace{-10pt}
	\caption[caption]{Version downgrade in the POODLE attack \cite{moller14}.} 
	\label{fig:poodle}
\end{figure}

In a typical downgrade attack, an active MitM attacker interferes with the protocol messages leading the communicating parties to operate in a mode weaker than they both support and prefer. Downgrade attacks have existed since the very early versions of TLS, SSL~v2 \cite{wagner96}. They can exploit various types of vulnerabilities (design, implementation, or trust-model), and target various elements of the protocol (algorithm, version, or layer) \cite{alashwali-taxonomy-18}. In the absence of handshake transcript authentication, downgrade attacks can be trivially performed. Starting from SSL~v3, the handshake transcript is authenticated at the end of the handshake to prevent downgrade attacks. However, experience has shown a series of downgrade attacks that circumvent the handshake transcript authentication. For example, \cite{beurdouche15}\cite{adrian15}\cite{moller14}\cite{beurdouche14}. \autoref{fig:poodle} shows version downgrade as in the POODLE \cite{moller14} attack.

\section{Preliminaries}
\label{sec:pre}
\subsection{Strict versus Default TLS Policy}
Our mechanism affects the client's fine-grained TLS configurations. Namely, the protocol version and ciphersuites. In addition, it affects the client's fallback mechanism. In our proposed mechanism, there are two pre-defined policies (or contexts) for the TLS client configurations: \textit{strict} and \textit{default}. The \textit{strict} policy enforces strong TLS configurations and disables the fallback. We define strong TLS configurations as those that support only the latest version of the protocol and only strong ciphersuites. We define strong ciphersuites as those that support both FS and AE properties simultaneously. The fallback is a mechanism that instructs the client to retry the handshake with weak configurations if the handshake with strong configurations has failed. On the other hand, the \textit{default} policy enforces both strong and weak TLS configurations, and enables the fallback. Weak configurations are defined as those that support both the latest and the legacy versions of the protocol, and both strong and weak ciphersuites. Weak ciphersuites are defined as those that support non-FS or non-AE. \autoref{table:modes} summarises the \textit{strict} versus \textit{default} policies that we define in our mechanism. Our prototypical TLS client implementation supports TLS versions: 1.0, 1.1, and 1.2, and 14 ciphersuites (similar to those supported in \texttt{Firefox} browser version 60.0.2 except that our client does not support the DES ciphersuite). Although TLS~1.3 is present, in our implementation and evaluation (\autoref{sec:implementation}) we consider TLS~1.2 as the latest TLS version. The reason is that TLS is currently in a transition state from version TLS~1.2 to TLS~1.3. TLS~1.3 has not been officially approved as a standard (is still a draft \cite{rescorla18tls13}), and is still in its beta version in most mainstream implementations such as \texttt{OpenSSL}. However, this does not affect our concept in general as it is applicable to the current deployment where TLS~1.2 is the latest version. Finally we note that in TLS~1.3, FS and AE ciphersuites are enforced by design \cite{rescorla18tls13}, i.e. strong ciphersuites are implied by TLS~1.3 as a version. Therefore, in TLS~1.3, the \textit{strict} configurations policy boils down to the protocol version and the fallback mechanism. However, there is still a value in our mechanism's ciphersuites policy even in TLS~1.3. Our policy enforces the client to refine its ciphersuites before the \texttt{ClientHello} is sent which provides downgrade resilience even when the server is flawed. This is unlike most TLS~1.3 clients, weak and strong ciphersuites are sent in the \texttt{ClientHello}, relying on the server to select the right version and ciphersuite. Experience shows that servers' flaws can be exploited to make the server select the wrong version as in \texttt{ClientHello} fragmentation \cite{beurdouche14}. 

\begin{table}[!tp]
	\caption{The \textit{strict} versus \textit{default} TLS policies that we define in our \name mechanism (\cmark denotes enabled and \xmark denotes disabled).}
	\label{table:modes}
	\centering
	\begin{tabular}{@{\extracolsep{4pt}}lllc@{}}
		\toprule
		Policy 				& TLS Version 							 & TLS Ciphersuites 			& Fallback \\
		\midrule
		\textit{Strict} 	& 	TLS~1.3   							 & FS and AE   					&\xmark	 \\
		\textit{Default} 	&	TLS~1.3; TLS~1.2; TLS~1.1; TLS~1.0	 & FS; AE; non-FS; non-AE		&\cmark	 \\ 
		\bottomrule
	\end{tabular}
	\vspace{-10pt}
\end{table}
 
\subsection{Problem Statement}
Achieving both security and backward compatibility is challenging. A \textit{strict} TLS client configurations policy provides stronger downgrade resilience than the \textit{default} one. However, the \textit{strict} policy may render many ordinary legacy servers unnecessarily unreachable, which results in a difficult user experience. On the other hand, the \textit{default} policy (such as mainstream web browsers today), provide backward compatibility but this is achieved at the cost of security. Experience shows that the \textit{default} policy can be abused by attackers to perform downgrade attacks as shown in the POODLE attack \cite{moller14}. \textbf{\textit{Can we achieve a better balance between the two extremes? Can we enable clients to enforce fine-grained TLS configurations based on prior knowledge about the servers' TLS configurations? Can we design a usable and authenticated mechanism that allows servers to advertise their support for strong TLS configurations so that clients can enforce a \textit{strict} TLS configurations policy for connections going to these servers while enforcing a \textit{default} configurations policy for the rest of the connections?}} 

\subsection{System and Threat Models}
Our system model considers the following parties: a TLS client, a TLS server, and a DNS server. A TLS server is identified by its domain name,  and the domain owner controls its DNS zone. These parties are standard for TLS connections and are assumed to be honest. As is the case of most real-world systems, the client and server support multiple protocol versions and ciphersuites that vary in the security guarantees they provide. Some of the versions and ciphersuites that the client and server support are weak, and are supported by both parties to be used \textit{if and only if} their peer is indeed a legacy one that does not support the strong configurations. The client and server aim to establish a TLS session using strong configurations. For example, if both parties support the latest version of the protocol (as of this writing, TLS~1.3), then both parties aim to use TLS~1.3. The DNS supports DNSSEC and uses strong signature algorithms and strong keys to sign the zone file which contains all the DNS records. The DNS keys are authenticated keys through a chain of trust in the DNS hierarchy. \par

In terms of threat model, we consider a MitM attacker who can passively eavesdrop on the transmitted messages, as well as actively modify, inject, drop, and replay messages during transmission. The attacker cannot break sufficiently strong cryptographic primitives (e.g. RSA signatures with 2048 bit ore more) that are properly deployed. The attacker does not have access to the DNS private-key that is used to sign the DNS zone file. We also assume the absence of MitM attackers in the first connection from the client to the DNS server for each domain. However, the MitM can exist in subsequent connections from the client to the DNS server.

\subsection{System Goals}
Our system goals can be summarised as follows:
\begin{compactitem}
	\item{Authentication:} TLS clients should be able to verify that the statement
	advertising the domain's support for the strong TLS configurations in the
	DNS is genuinely produced by the domain owner.
	\item{Usability:}
	The mechanism should be usable to the clients' end users. It should not incur additional manual configurations on the users.
	\item{Compatibility:}
	The mechanism should be compatible with existing Internet infrastructure. It should not require additional infrastructure or trusted third parties above those in a typical TLS connection. 
	\item{Performance:}
	The mechanism should be lightweight. It should incur minimal overhead on the clients' performance.
\end{compactitem}

\section{The \name Mechanism}
\label{sec:overview}
\subsection{Overview}
Our mechanism aims to provide a usable and authenticated method that allows domain owners to advertise their support for strong TLS configurations to TLS clients. This provides the clients with prior knowledge that enables them to take an informed decision on whether to enforce a \textit{strict} or \textit{default} TLS configurations policy, before connecting to a domain. Throughout the paper, we refer to the \name record in the DNS as the \name policy record. 

\subsection{\name Policy Syntax} \label{sec:policy-syntax}
In what follows, we describe each directive used in the \name policy syntax. \autoref{fig:format} shows an example of an ideal \name record in a DNS \texttt{zone} file.
\begin{compactitem}
\item \texttt{name:} Specifies an identifier for the \name records. Our mechanism uses a general purpose DNS record (\texttt{TXT}). Therefore, the record must be identified as a \name to be interpreted by clients as a \name policy record. This directive value must be set to \texttt{\name}.
\item \texttt{validFrom:} Specifies the \name policy issuance date. It indicates the recency of the policy. It acts as a version number for the policy when there are multiple issued policies. The most recent must be the effective one. This directive value takes a date in a \texttt{dd-mm-yyy} format.
\item \texttt{validTo:} Specifies the \name policy expiry date. It indicates the validity of the policy. This directive value takes a date in a \texttt{dd-mm-yyy} format.
\item \texttt{tlsLevel:} Specifies the TLS level that the server advertises. This directive value must be set to \texttt{strict-config} for the \textit{strict} TLS configurations policy to be enforced by the client.
\item \texttt{includeSubDomain:} Specifies whether the policy should be enforced to sub-domains or not. It takes either \texttt{0} to disable the option or \texttt{1} to enable it.
\item \texttt{revoke:} Specifies whether the domain wants to opt-out from the \name policy or not. It takes either \texttt{0} to disable the option or \texttt{1} to enable it. If enabled, it acts as a poisoning flag. When a server wants to opt-out from the \name, it should keep advertising a \texttt{revoke} with value \texttt{1} until the expiry date of any previously published \name policy. This instructs clients to delete the revoked \name from their storage if exists.
\item \texttt{report:} Specifies the email address of the domain owner. It takes a string in an email address format. The email can be used by TLS clients to allow the user to report a domain's failure of complying with the advertised policy to the domain owner.
\end{compactitem}

\subsection{Details}
\begin{figure*}[tp!]
	\centering
	\includegraphics[width=\textwidth]{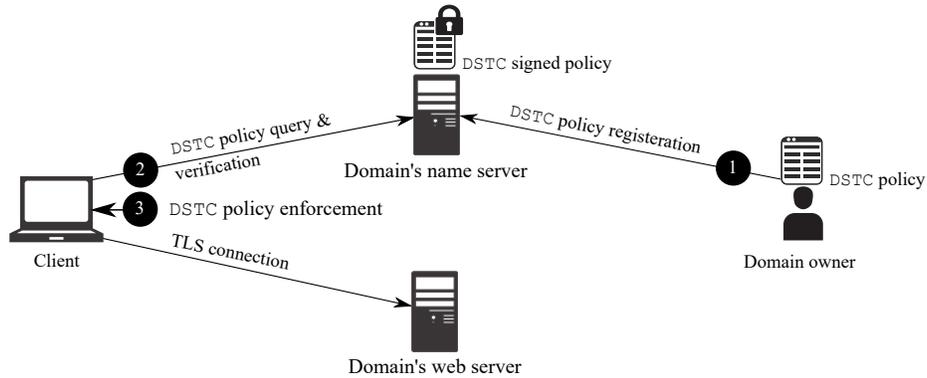}
	\caption{A high-level overview of the DSTC mechanism.}
	\label{fig:overview}
\end{figure*}

The mechanism can be summarised in three main phases as follows: 

\begin{compactenum}
\item \textbf{Policy Registration:} 
	\begin{compactenum}
		\item The policy must be defined by the domain owner according to the policy syntax in section \autoref{sec:policy-syntax}. 
		\item The policy needs to be published as a \texttt{TXT} record in the DNS by the domain owner.  
		\item The policy needs to be signed by the domain owner using the private-key of the \texttt{ZSK}. By the end of this step, the signed \name policy is publicized in the DNS in the domain's \texttt{TXT} record.    
	\end{compactenum}
\vspace{10pt}

\begin{figure}[tp!]
	\centering
	\setlength{\fboxsep}{3pt} 
	\setlength{\fboxrule}{0.5pt} 
	\fbox{\includegraphics[width=\textwidth]{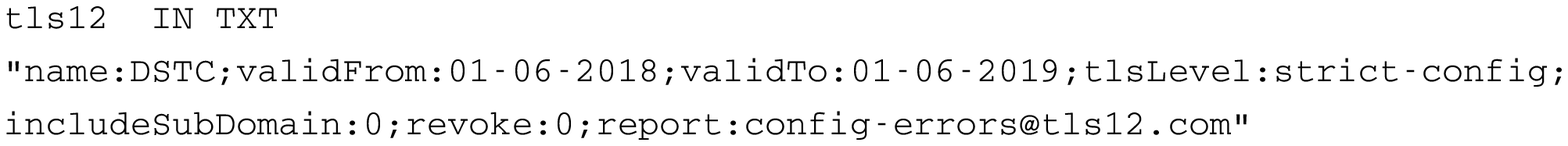}}
	\caption{An example of a \name record in the DNS for the domain \quotes{tls12}.}
	\label{fig:format}
\end{figure}

\item \textbf{Policy Query and Verification:} 
	\begin{compactenum}
		\item When a client wants to connect to a website, the client queries the DNS to retrieve the domain's DNS records. The \name is returned in a signed \texttt{TXT} record. 
		\item The client verifies the signature using an authenticated public-key of the \texttt{ZSK}. If the signature is valid, the client verifies the rest of the \name policy directives. Based on the verification result, this step returns a value that signals the TLS configuration policy to be enforced: either \textit{strict} or \textit{default} along with a message to clarify the status (e.g. invalid signature) and the reporting email. The \textit{strict} policy is returned only when all the verifications pass. Otherwise, the policy remains \textit{default}.    
   	\end{compactenum}
\vspace{10pt}
\item \textbf{Policy Enforcement:} 
	\begin{compactenum}
		\item The client receives the TLS configuration policy from the previous step (Query and Verification). 
		\item The client enforces the policy according to the policy received: either \textit{strict} or \textit{default}. 
	\end{compactenum}
\end{compactenum} 
\vspace{10pt}
After the TLS configurations policy is enforced, which affects the TLS \texttt{\justify{ClientHello}} offered versions and ciphersuites parameters, the client connects to the server. \autoref{fig:overview} illustrates the \name system and the actors involved. The TLS connection is not part of our mechanism phases, but we include it in \autoref{fig:overview} to provide a complete view of the system. 
 
\section{Security Analysis}
\label{sec:analysis}
In our system, the attacker wins under two conditions: First, if he can forge a \name policy and
present it to a \name-supported TLS client as a valid policy. Second, if he can perform an undetectable TLS version or ciphersuite downgrade attack that makes a \name-supported TLS client accept weak TLS configurations despite the downgrade-resilience that the \name policy provides.

\subsection{\name Forgery}
An active MitM attacker can achieve \name forgery if he can add, modify, delete, drop, or replay a \name policy record for a particular domain. The attacker's gain from each method can be summarised as follows:
First, adding a policy for a domain that did not register a \name policy can cause a Denial of Service
(DoS) attack for that domain. When \name-supported clients enforce a \textit{strict} configurations policy for a domain that actually did not register a \name record and does not comply with the policy's requirements (e.g. uses a legacy protocol version), this will result in aborted handshake by the client. Second, modifying a \name policy record's directives can cause either DoS or Denial of Policy (DoP) for the concerned domain, depending on the modified directive. DoP prevents a policy from being enforced despite the domain's registration, which results in \textit{default} client configurations which in turn provides weaker downgrade-resilience than desired. For example, modifying the \texttt{validTo} directive to an earlier date than it actually is, results in DoP since the policy will be marked as expired by the client at some point of time, and will not be enforced, while it is expected to be enforced by the domain. On the other hand, modifying the \texttt{validTo} directive to a later date results in DoS since the policy will be enforced for a domain that is not advertising the policy and may no longer complying with it. Third, deleting a \name policy record will result in DoP since the client does not get the \name record and enforces the \textit{default} TLS configurations, which provides weaker downgrade-resilience. Fourth, replaying a non recent or revoked policy that has a valid signature can cause a DoS or DoP attacks as explained above. \par

In our system, adding, modifying, or deleting a \name policy record for a domain is defeated by the digital signature. The DNSSEC is a mandatory component of the system where \name records are signed by the domain owner using the private-key of the \texttt{ZSK}. The attacker does not have access to the DNS private-key and does not have the power to break it or break the signature algorithm. Regarding replay attacks, the client stores the policy locally and updates or revokes (deletes) it when a signed, more recent (i.e. more recent \texttt{validFrom} date), and non-expired policy is received. A replayed outdated or revoked policy will have a less recent issuance date than the stored one, and hence will be detected even if it has a valid signature. Finally, dropping attacks are also defeated by the stored policy from the first connection which is received under the assumption of the absence of MitM in the first connection from client to DNS. If the client has a non-expired stored policy, and the client has not received any new \texttt{revoke}-enabled policy to instruct the client to delete it, the absence of the \name record in subsequent DNS queries signals a \name dropping attack. Note that connections after the stored \name policy expires are considered a first connection and assumed to be in a MitM-free connection. \par

\subsection{TLS Downgrade Attacks}\label{downgrade-attack}
We now show how the \name mechanism prevents a class of downgrade attacks that
abuse the client's support for legacy configurations and silent fallback. We
demonstrate it on real-world downgrade attack scenarios.\par

\begin{figure}[!t]
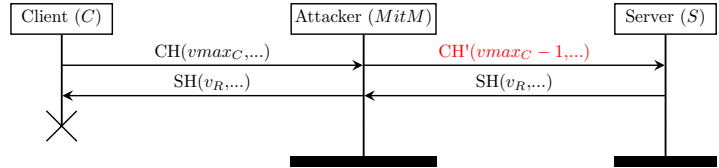
 
	\centering
	\resizebox{0.80\columnwidth}{!}{
		\setmsckeyword{} 
		\drawframe{no} 
			\begin{msc}[large values, /msc/level height=0.6cm, /msc/label distance=0.5ex , /msc/first level height=0.6cm, /msc/last level height=0.6cm, /msc/top head dist=0, /msc/bottom foot dist=0, /msc/environment distance=0, /msc/foot height=1.5ex]{}
			\setlength{\instwidth}{2\mscunit} 
			\setlength{\instdist}{3.5\mscunit} 
			
			\declinst{I}{}{Client ($C$)}%
			\declinst{M}{}{Attacker ($MitM$)}%
			\declinst{R}{}{Server ($S$)}%
			
			\mess {CH($vmax_{C}$,...)}{I}{M}
			\mess {\textcolor{red}{CH\textquotesingle{}($vmax_{C}-1$,...)}}{M}{R}
			\nextlevel
			
			\mess {SH($v_{R}$,...)}{R}{M}
			\mess {SH($v_{R}$,...)}{M}{I}
			\nextlevel
			\stop{I}
		\end{msc}
	} 
	\vspace{-10pt}
	\caption[caption]{Illustration of a version downgrade attack with a \name-supported client.} 
	\label{fig:key-exchange downgrade}
\end{figure}

The first scenario is inspired by the \texttt{ClientHello} fragmentation version downgrade attack~\cite{beurdouche14}. In this attack, due to a flawed TLS server implementation,
if an attacker fragments the \texttt{ClientHello}, the server falls back to TLS~1.0. A \textit{default} client
will silently fall back to TLS~1.0 under the assumption that it is connecting to a legacy
server. However, with a \name-supported client and registered server, this attack is defeated as the client enforces a \textit{strict} TLS policy and does not fallback, hence the attack will be detected and the handshake will be aborted. \par

The second scenario is inspired by the POODLE version downgrade attack \cite{moller14}. In this attack, the attacker drops the \texttt{ClientHello} message one or more times. Some TLS clients interpret this as a server compatibility issue and retry to send the \texttt{ClientHello} using a lower version. With a \name-supported client, the client does not fallback since it has prior knowledge about the server's support for strong configurations, hence the attack will be detected and the handshake will be aborted.

\section{Implementation and Evaluation}
\label{sec:implementation}
\subsection{Applicability}\label{applicability}
To get an insight into the applicability of our proposed mechanism,
we conduct a TLS scan (IPv4 space) for the top 10,000 most visited Internet domains globally.
The scan provides quantitative data about the supported and preferred TLS
versions and ciphersuites in real-world servers. We retrieve the top 10,000 domains
list\footnote{The list gets updated daily, according to Alexa's support (in a private communication).} 
from Alexa Internet \cite{alexa18} on the 5\textsuperscript{th} of May
2018. To run the scan, we use \texttt{sslscan 1.11.11} \cite{sslscan18}, a state-of-the-art open source TLS scanning tool that can perform TLS versions and ciphersuites enumeration through multiple TLS handshakes. The tool supports SSLv2 up to TLS~1.2, and 175 ciphersuites. We run the scan from the SUTD university's campus wired network between the 6\textsuperscript{th} and 12\textsuperscript{th} of May 2018. In terms of ethical considerations, our scan does not collect any private or personal data. The TLS versions and ciphersuites are public data which can be viewed by TLS clients through TLS handshakes. The number of handshakes the tool performs does not represent a danger of DoS. \par

The total number of servers that completed a successful TLS handshake with one or more TLS versions and ciphersuites is 7080 (70.80\%). We do not investigate the reasons of handshake failure as this is outside our scope. However, a recent study that performed domain name-based TLS scans for various domains \cite{amann17}, reports 55.7 million and 58.0 million successful TLS handshakes out of 192.9 million input domains (29.48\% on average). Given the fact that our scan is for top domains, our TLS response rate sounds normal. However, one possible contributing factor to the handshake failure in our scan can be due to SUTD university's Internet censorship system that blocks some website categories such as porn and gambling. \par 

In terms of TLS versions, of the responding servers in our results, there are 6888 (97.29\%) servers that support TLS~1.2. TLS~1.2 is the preferred version in all the servers that support it. However, there are only 373 (5.27\%) servers that support TLS~1.2 exclusively (without any other versions). On the other hand, the number of servers that support at least two version, both TLS~1.2 and TLS~1.1, either exclusively or with other lower versions, is 6462 (91.27\%). And the number of servers that support at least three versions, TLS~1.2, TLS~1.1 and TLS~1.0, either exclusively or with other lower versions, is 6202 (87.60\%). \par 

In terms of ciphersuites, we examine the servers' ciphersuites in version TLS~1.2 only. The most frequent number of supported ciphersuites (the norm) is 20 ciphersuite, which appeared in 938 servers (13.62\%). To count the servers that support FS and/or AE, in each domain in our results, we labeled each supported ciphersuite by one of the following labels: FS+AE, FS+nonAE, nonFS+AE, or nonFS+nonAE. The four labels are based on the two properties: FS and AE. FS is identified by checking if the ciphersuite starts with \texttt{ECDHE} or \texttt{DHE}, while AE is identified by checking if the ciphersuite contains \texttt{GCM}, \texttt{CCM}, \texttt{CCM8}, or \texttt{ChaCha20} strings. There are 6500 (94.37\%) TLS~1.2 servers containing at least one FS+AE ciphersuite, either exclusively or with other labels. We find 6483 (94.12\%) TLS~1.2 servers that support non-FS or non-AE (i.e. labeled with nonFS+AE, FS+nonAE, or nonFS+nonAE) in addition to one or more FS+AE ciphersuite. \par

The results show that top domain servers support the strong TLS configurations. At the same time, they maintain support for weak configurations that have known weaknesses and provide fewer security guarantees. Ideally, the clients' configurations influence the servers' selected configurations. Asserting servers' strong configurations to clients adds a value by providing clients with the confidence to enforce a \textit {strict} TLS configurations policy for connections to these servers, which reduces the downgrade attack surface as we showed in \autoref{downgrade-attack}.  
     
\subsection{Feasibility} \label{feasability}
To test the feasibility of our concept, we implement a Proof-of-Concept (PoC) for the mechanism. On a machine equipped with 16 GB Random Access Memory (RAM) and \texttt{Intel Core i7} 2.6 GHz processor, and runs \texttt{Windows 10} (64-bit) OS, we build a virtual private network with a virtual host-only Ethernet adapter using \texttt{VirtualBox} \cite{virtualbox18}. It includes four virtual machines: Three TLS web servers, a DNS server, and a TLS client. The web servers are equipped with 2 GB of RAM, \texttt{Intel Core i7} CPU 2.60 GHz processor, and 1000 Mbps wired network card. They run \texttt{Apache 2.4.18} \cite{apache18} on \texttt{Ubuntu 16.04} (64-bit) Operating System (OS). The DNS server is similar to the web servers in specifications except that it has 4 GB RAM and runs \texttt{BIND 9.10.3} \cite{bind}. The DNS server supports DNSSEC and the zone file is signed with a 2048 RSA \texttt{ZSK}. The \texttt{ZSK} is signed with a 2048 RSA \texttt{KSK}. We assume the \texttt{KSK} is validated through a chain of trust. To evaluate a \name-supported client, we implement a TLS client using \texttt{Python 3.6.5} \cite{python} and \texttt{python}'s TLS/SSL library \cite{pythonssl} on a Linux \texttt{Ubuntu 18.04} (64-bit) OS on a device equipped with 4 GB of RAM, \texttt{Intel Core i7} CPU 2.60 GHz processor, and 1000 Mbps wired network card. The client uses \texttt{OpenSSL 1.1.0g} that is shipped with \texttt{Ubuntu 18.04}. In our PoC we assume the highest version of TLS is TLS~1.2. Therefore a \name-compliant server should comply to TLS~1.2 and strong ciphersuites. Our client initiates a handshake with the three TLS web servers. The servers are configured as follows: First, to represent a \name compliant server that has registered a \name record, we configure a TLS~1.2 server with strong ciphersuites, and register a \name policy record for it in the DNS. Second, to represent a downgrade attack or misconfigured server, we use a straight-forward method to make the server's version lower than the \name requirements, we configure a TLS~1.0 server and add a \name policy record for it. Third, to represent a server that has not registered a \name record which should not be affected, we configure a TLS~1.1 which does not comply with the \name requirements and we do not register a \name policy record for it. \par 

As depicted in \autoref{table:testing}, the handshake with the first server succeeds as the server complies with the \name requirements. The handshake with the second server fails as the server fails to comply with the \name requirements. The handshake succeeds with the third server as the server did not register a \name policy record. Our experiment confirms that the concept is technically feasible. 

\begin{table}[!tp]
  \caption{Test-case scenarios carried from our \texttt{python} DSTC-supported client to TLS servers and the effect of DSTC (\cmark denotes \name registered domain and \xmark denotes unregistered) on the TLS handshake (\cmark denotes successful and \xmark denotes failed).}
  \label{table:testing}
  \centering
  \begin{tabular}{@{\extracolsep{4pt}}lllcc@{}}
    \toprule
    \multicolumn{1}{c}{\multirow{2}{*}{No.}} & \multicolumn{3}{c}{TLS Server Configurations}  & \multicolumn{1}{c}{\multirow{2}{*}{\makecell{Successful\\ Handshake}}}\\
    \cline{2-4}
      & 	Version 		& Ciphersuites Feature 		& DSTC 		&  		\\
    \midrule
    1 & 	TLS~1.2			& FS and AE   				&\cmark		&\cmark \\
    2 & 	TLS~1.0			& non-AE 					&\cmark		&\xmark \\
    3 & 	TLS~1.1			& non-AE					&\xmark		&\cmark \\
  	\bottomrule
  \end{tabular}
\vspace{-10pt}
\end{table}

\subsection{Performance}\label{performance}
To get an insight into the computational cost that our mechanism adds over an ordinary TLS connection, based on scenario 1 in \autoref{table:testing} (assuming no cached policy in the client) we measure the execution time for the following functions: \texttt{SigVerify} for the DNS \texttt{TXTRRset} records signature verification, \texttt{QueryVerify} for the DNS records query and verification (which includes \texttt{SigVerify}), \texttt{Enforce} for the TLS policy enforcement based on the \texttt{QueryVerify} output, and finally, the time for the three functions together. \autoref{table:performance} presents the measurements using the processor timer in \texttt{python}'s \texttt{3.6  time} module \cite{pythontime18}, which is processor-wide timer. Each measurement is repeated 500 times. A TLS socket connection establishment in our client takes 8.16 ms on average (without certificate validation). The mechanism's overall average overhead costs 3.58 ms. We conclude that the computational overhead is affordable which is about 43.87\% additional overhead on the TLS socket connection. Our mechanism's overhead can be considered an upper-bound as there is a room for improvements through code optimisation.

\begin{table}[!tp]
	\caption{The mechanism's computational overhead in milliseconds.}
	\label{table:performance}
	\centering
	\begin{tabular}{@{\extracolsep{4pt}}llllllll@{}}
		\toprule
		No. & Function 					& Max.  		& Min.  				& Avg. \\
		\midrule
		1 & \texttt{SigVerify} 			& 1.40			& 0.63					& 0.72					\\
		2 & \texttt{QueryVerify} 		& 4.99			& 2.74					& 3.09					\\
		3 & \texttt{Enforce} 			& 0.86			& 0.38					& 0.41					\\
		4 & \texttt{All 3 functions} 	& 6.10			& 3.23					& 3.58					\\
		\bottomrule
	\end{tabular}
	\vspace{-10pt}
\end{table}

\section{Related work}
\label{sec:related}
Schechter~\cite{schechter07} proposes the HTTP Security Requirements in the Domain Name System (HTTPSSR DNS). It allows domain owners to assert their support for the TLS protocol to prevent TLS layer downgrade (a.k.a. stripping) attacks. However, experience shows that asserting TLS (as a layer only) is not sufficient. Several downgrade attacks that target TLS configurations such as the protocol version or ciphersuite as in the POODLE version downgrade \cite{moller14} have been shown successful. Dukhovni and Hardaker~\cite{dukhovni15} propose the DNS-based Authentication of Named Entities (DANE). It allows domain owners to bind their own CA public keys or certificates to detect faked TLS certificates to prevent domain impersonation attacks. Hallam-Baker~\cite{hallam13} proposes the Certificate Authority Authorisation (CAA). It allows domain owners to whitelist specific Certificate Authorities (CAs) for their domains to prevent mis-issued certificates. Alashwali and Rasmusssen~\cite{alashwali18}  propose client \textit{strict} TLS configurations against whitelisted domains as a downgrade attacks defense. The domains are added either by the client's users or through servers' HTTP headers. While adding domains through the servers' headers is usable, the \textit{strict} policy can only be enforced starting from the second connections (the first connection is configured before the headers are fetched and hence uses \textit{default} configurations). Our scheme extends this work by leveraging DNS which allows the \textit{strict} policy enforcement before the first connection in a usable and authenticated manner without extra effort from clients' users. Finally, Varshney and Szalachowski~\cite{varshney18} propose a general DNS-based meta-policy framework. Overall, none of the previous work have looked at using DNS to enable domain owners to assert strong TLS configurations.

\section{Conclusion}
\label{sec:conclusion}
We propose a mechanism that allows domain owners to advertise their support for strong TLS configurations through a signed DNS record. The client interprets this record and changes its behaviour to the \textit{strict} policy which affects the TLS version, ciphersuite, and the fallback mechanism. Our prototype implementation and its evaluation show the feasibility of our mechanism. Furthermore, our Internet scan results depict that the majority of servers are ready to benefit from the proposed mechanism.

\section*{Acknowledgement}
We thank Prof. Andrew Martin for feedback and Monica Kaminska for proofreading. Pawel's work was supported by the SUTD SRG ISTD 2017 128 grant.

\bibliographystyle{splncs03}
\bibliography{ref}
\clearpage
\end{document}